\documentclass{aastex}

\usepackage{emulateapj5}

\input epsf
 
\def\ion#1#2{#1\,{\sc #2}}
\newcommand{\capella} {\emph{Capella}}
\newcommand{\gal} {$\alpha$}
\newcommand{\gb} {$\beta$}
\newcommand{\ggam}{$\gamma$}

\newcommand{\gl} {$\lambda$}

\newcommand{\etal} {et al.}  
\newcommand{\ecs} {erg\,cm$^{-2}$\,s$^{-1}$} 
\newcommand{\kms} {km\,s$^{-1}$} 
\newcommand{\deltE}{\Delta\kern-1ptE}
\newcommand{\fuse}{\emph{FUSE}}

\newcommand{\lam}{$\lambda$}

\begin{document}

\title{FUSE Observations of Capella}
 
\author{P. R. Young,\altaffilmark{1}
A. K. Dupree,\altaffilmark{1}
B. E. Wood,\altaffilmark{2}
S. Redfield,\altaffilmark{2}
J. L. Linsky,\altaffilmark{2}
T. B. Ake,\altaffilmark{3,4}
\and H. W. Moos\altaffilmark{3}}

\altaffiltext{1}{Harvard-Smithsonian Center for Astrophysics, 60
Garden Street, Cambridge, MA 02138}

\altaffiltext{2}{JILA, University of Colorado and NIST, Boulder, CO
80309-0440}

\altaffiltext{3}{Department of Physics and Astronomy, Johns Hopkins University,
3400 North Charles Street, Baltimore, MD 21218}

\altaffiltext{4}{Computer Science Corporation, Lanham, MD 20706}

\begin{abstract}

\emph{Far Ultraviolet Spectroscopic Explorer} (\fuse) observations
of the bright binary system \capella\
(\gal\ Aurigae; G1 III + G8 III) reveal a rich emission line spectrum
containing neutral and ionic species, among them \ion{H}{i}, \ion{O}{i},
\ion{C}{iii}, \ion{O}{vi}, \ion{S}{vi}, \ion{Ne}{v} and
\ion{Ne}{vi}. In addition the \ion{Fe}{xviii} 974.85\,\AA\ line,
formed at temperatures of $\approx$\,$6\times 10^6$~K, is
detected. Whereas the chromospheric and transition region emission is
dominated by that from the G1
giant, consistent with results from previous ultraviolet observations,
\ion{Fe}{xviii} is formed largely in the G8 giant atmosphere.  Line
ratios from \ion{C}{iii} suggest densities of 2--8$\times$10$^{10}$
cm$^{-3}$, although anomalous line profiles of the $\lambda$1176
transition may signal optical depth effects.  The hydrogen Lyman
series, detected for the first time, displays asymmetric emission
consistent with an expanding atmosphere.

\end{abstract}


\section{Introduction}

The \capella\ system ($\alpha$ Aurigae = HR 1708 = HD 34029), consisting
principally of two nearby cool giant stars (G1~III + G8~III) forming a
spectroscopic binary, has long been a popular object because it is a
bright target at virtually all energies.  In the far ultraviolet
(900--1200\,\AA), \capella\  was first observed with {\it Copernicus}
in H Ly\gal\ and \ion{O}{vi} \gl1032  (Dupree 1975), and subsequently
with HUT (Kruk \etal\ 1999) and the Orbiting Retrievable Far and Extreme
Ultraviolet Spectrometers (\emph{ORFEUS I}).
High resolution data with the \emph{International Ultraviolet Explorer} (\emph{IUE}; Ayres~1988)
and the \emph{Hubble Space Telescope} Goddard High Resolution
Spectrograph (\emph{HST}/GHRS; Linsky et al.~1995) spectroscopically
resolves the two giants through velocity separation.
Direct UV image
separation of the chromospheric and transition region emissions has
also been achieved with \emph{HST} (Young \& Dupree 2001).  \capella\
is a target of the PI team  on \fuse,   forming a part of the cool
star survey and local D/H program.

\section{FUSE Observations and Spectra}

\fuse\ is described in detail by Moos \etal\ (2000).  Four individual
telescopes  cover  the wavelength range 905--1187\,\AA\ and they  are
denoted by the coatings on their optical elements: LiF1, LiF2, SiC1
and SiC2.  Spectra from each channel are imaged onto two detector
segments, labelled A and B so that the eight individual spectra are
termed  LiF1A, LiF1B, SiC1A, etc.  Three apertures are available for
observing: in order of decreasing size these are the LWRS, the MDRS
and the HIRS. For the two smaller apertures, there is a risk of losing
the target in the aperture during the observation, but they have the
advantage of lower airglow contamination compared to the LWRS and so
are preferred for, e.g., the study of the hydrogen Lyman series lines.

Within a single spectrum, the relative wavelength calibration is good
to 5--10~km/s, but the absolute wavelength calibration of \fuse\ is
not well determined as there is no onboard calibration lamp. The
wavelength scale can be  fixed, however, by making use of interstellar
absorption lines which have a velocity known from other methods. For
emission line sources such as cool stars, the lack of a continuum
means that such absorption lines are not common. The two most
commonly observed are \ion{C}{iii} 977.020\,\AA\ and \ion{C}{ii}
1036.337\,\AA\ which are superimposed upon the corresponding stellar
emission lines. They potentially allow the wavelength scale to be set
for all channels except LiF2A and LiF1B. Caution must be exercised,
however, as the \lam977 absorption may be intrinsic to the star in
some cases.

\capella\ was observed on  5 and 7 Nov.\ 2000 with the LWRS aperture
(program IDs P1041301 and P1041302) for 14.2 and 12.3~ksec.  A further
observation of 21.2~ksec was 
obtained on  11 Jan.\ 2001 in the MDRS aperture (program ID
P1041303).  A loss of flux occurred in all channels in the latter
observation as the target moved in  the aperture making emission line
fluxes unreliable.  The data were processed with v.1.8.7 of the \fuse\
calibration pipeline.  The ephemeris of Hummel \etal\ (1994) and the
radial velocity measurements of Barlow \etal\ (1993) indicate that the
heliocentric radial velocities of  the stars during the three
observations are: +7.1, +8.1, +32.8~\kms\ (G8 giant), and +52.4, +51.3
and +25.4~\kms\ (G1 giant).

\begin{figure*}[h]
\figurenum{1}
\begin{center}
\vbox{
	\epsfxsize=18cm\epsfbox{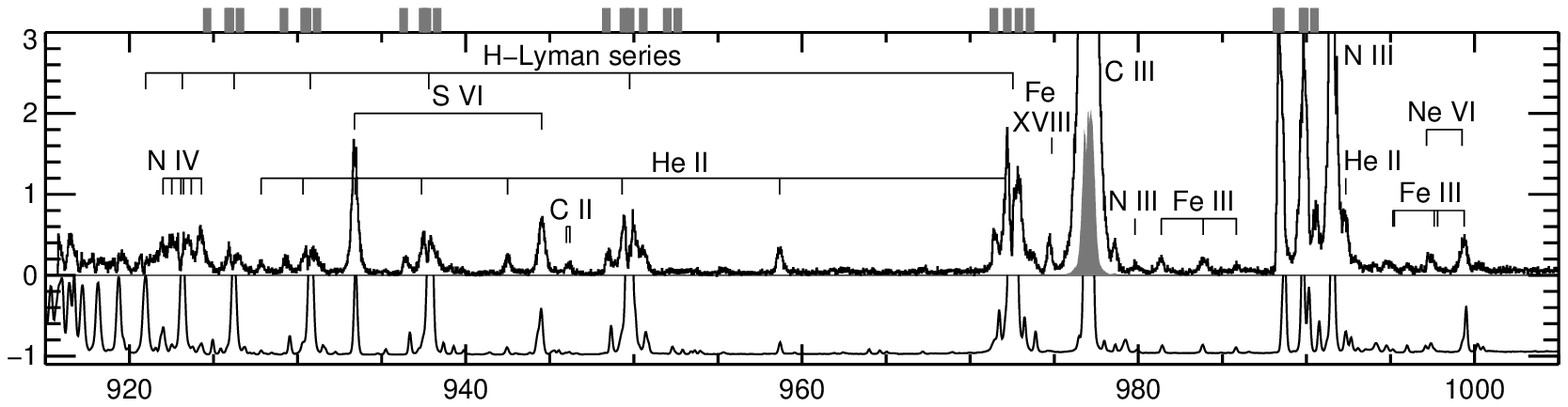}
	\epsfxsize=18cm\epsfbox{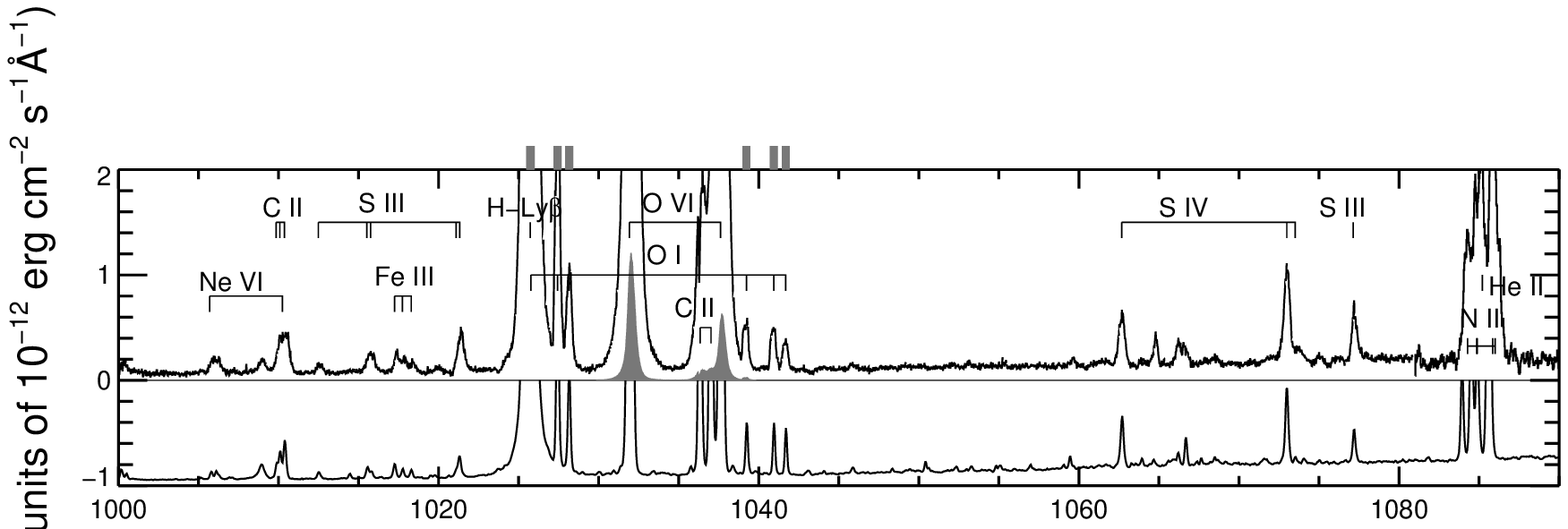}
	\epsfxsize=18cm\epsfbox{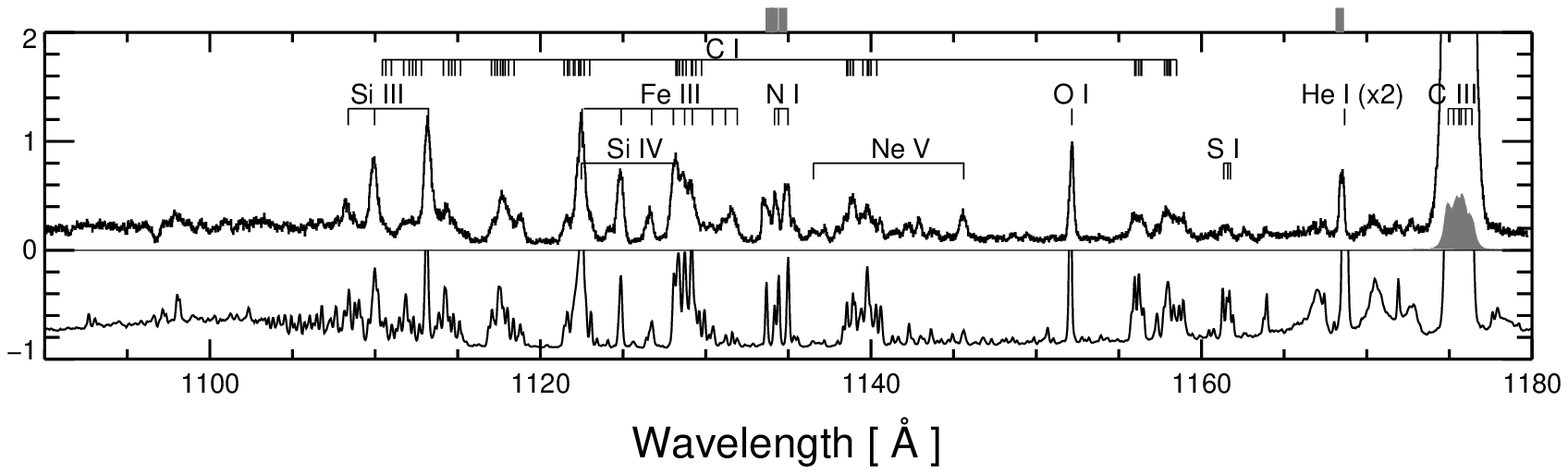}
}	
\end{center}

\caption{The LWRS \fuse\ spectrum of \capella\ derived from  observation
P1041301. The quiet 
solar spectrum shown below in each panel was obtained by the
SUMER instrument on SOHO (Curdt \etal\ 1997; 
W.\ Curdt, private communication). The shaded regions above each
spectrum denote areas contaminated by airglow lines. The \capella\
\ion{C}{iii} \lam977, \lam1176, and \ion{O}{vi} \lam1032, \lam1038
lines, reduced by a factor 18, are plotted as shaded profiles.}
\label{fuse-spec}
\end{figure*}

Fig.~\ref{fuse-spec} shows spectra from observation P1041301 that
cover the entire \fuse\ spectral range together with a solar spectrum.
Identifications are based on the solar spectrum and  predictions from
CHIANTI (Dere \etal\ 2001) using the emission measure distribution and
densities from UV and EUV observations of \capella\ (Brickhouse et
al.\ 2000).  The \capella\ spectrum is strikingly similar to the solar
spectrum both in terms of the relative  strengths of the stronger
lines and the strengths of the numerous weaker features. The central
differences lie in the broader line profiles and the presence of
\ion{Fe}{xviii} \lam974 (discussed later) in \capella.  Some lines are
proportionately weaker or stronger in the \emph{Capella} spectrum, for
example the \ion{N}{iv} lines at 922--924\,\AA\ are stronger which may
indicate abundance anomalies in the G8 giant as suggested in Linsky et
al.~(1995).  The broader lines found in \emph{Capella} are due both to
the combined emission from the two  giants and the intrinsically
larger broadening of the G1 lines.  We draw attention to the
\ion{C}{i} recombination continuum extending  below $\approx
1100$\,\AA\ to $\approx 1000$\,\AA\ which is seen clearly in both
spectra. Several absorption lines can be seen superimposed on this
continuum and these, together with the interstellar absorption lines,
will be discussed in a later paper (S.\ Redfield et al., in
preparation).

Table 1 presents fluxes and total counts in selected emission
lines. A complete line list for the spectrum is available at
\url{http://fuse.pha.jhu.edu/analysis/cool\_stars/capella\_linelist.html}.
Although dominated by the resonance lines of \ion{C}{iii} and
\ion{O}{vi}, the spectrum exhibits a wealth of weaker lines.
Elements 
represented include H, He, C, N, O, Ne, Si, S and Fe. We discuss
features of individual lines below.

\section{O VI Emission}

The \ion{O}{vi} \lam1032 line is unblended and displays broad wings
that extend out to $\approx$\,$\pm 600$ \kms\ from line center, which
could represent microflaring activity (Wood \etal\ 1997) or emission
from regions extending well above the stars' surfaces. Use of the
\ion{C}{ii} \gl1036 absorption line to fix the wavelength of the
\gl1032 line shows that the G1 star dominates the \gl1032 emission,
consistent with lower temperature transition region lines
observed with \emph{IUE} and GHRS (Ayres 1988; Linsky \etal\
1995). The \ion{O}{vi} \lam1038 line is partially blended with the
\ion{C}{ii} 1037.018\,\AA\ emission line, but the \gl1038/\gl1032
ratio is $\approx$\,1:2, indicating that the plasma is optically thin
at these wavelengths.

\section{C III and Chromospheric Density}

The ratio of the \ion{C}{iii} \gl977 and \gl1176 lines provides an
excellent electron density diagnostic between $10^8$ and
$10^{11}$~cm$^{-3}$ for optically  thin plasmas (Dupree \etal\ 1976).
The observed flux ratio (erg units)  \gl1176/\gl977 from observation
P1041301 equals
0.51$\pm$0.01 which translates to an electron density of
2$\times$10$^{10}$ cm$^{-3}$ using atomic data from CHIANTI. The drop
in the \lam977 flux in the second LWRS observation (Sect.~7) leads to
a higher density of 8$\times$10$^{10}$ cm$^{-3}$.
The
corresponding electron pressures are 
$\approx$\,2--8$\times 10^{15}$~K\,cm$^{-3}$, values consistent with the
\ion{C}{v} pressure 
obtained by Ness \etal\ (2001) from \emph{Chandra} spectra  and would
indicate constant pressure through the \emph{Capella} giants' transition
regions.  Much higher pressures are found at higher temperatures, based
on EUV density diagnostics of \ion{Fe}{xix}
and \ion{Fe}{xxi} which yield densities of $\sim$10$^{12}$~cm$^{-3}$
(Dupree \etal\ 1993; Brickhouse 1996), indicating inhomogeneous
structures.

\begin{figure*}[h]
\epsscale{.5}
\figurenum{2}
\plotone{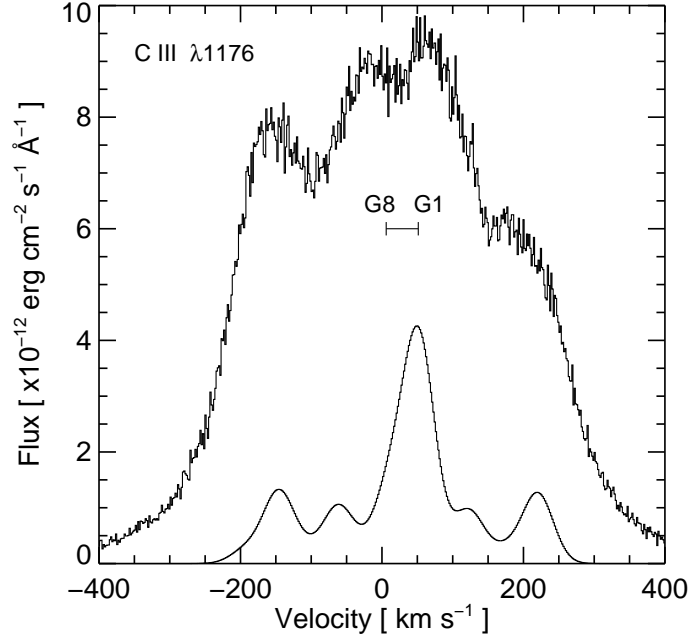}
\caption{A comparison of the observed \ion{C}{iii} \lam1176 profile (upper)
from the LWRS observation P1041301 with a model profile (lower) derived with CHIANTI for an
electron density of
$10^{10}$~cm$^{-3}$.}
\label{capella-1176}
\end{figure*}

The derived density from the \ion{C}{iii} lines must be treated with
caution as they show
evidence of optical depth effects. Fig.~\ref{capella-1176} compares
the \capella\ \gl1176 profile with a simulated optically thin profile
calculated using CHIANTI and assuming a cool-to-hot giant ratio of
1/5. The \gl1176 feature comprises six individual transitions spread
over 1.5\,\AA\ and although the lower levels of the transitions belong
to an excited configuration, they can have populations comparable to
the ground $^1$S level at electron densities $\gtrsim
10^{10}$~cm$^{-3}$, allowing the possibility of photoabsorption from
these levels.  Clearly the strong peak expected from the
1175.711\,\AA\ transition (the strongest of the multiplet) is much
weaker in \capella.  This reduction in the observed strength of the
1175.711\,\AA\ line compared to the optically thin case has been
observed in solar spectra  taken near the
limb which indicate  that the \gl1175.711/\gl1174.933 ratio can fall
to 50\% of the optically thin value (Doyle \& McWhirter 1980).
Further work is required to
model radiative transfer in the \ion{C}{iii} lines.

\section{The Lyman Series of Hydrogen}

\begin{figure*}[h]
\epsscale{0.5}
\figurenum{3}
\plotone{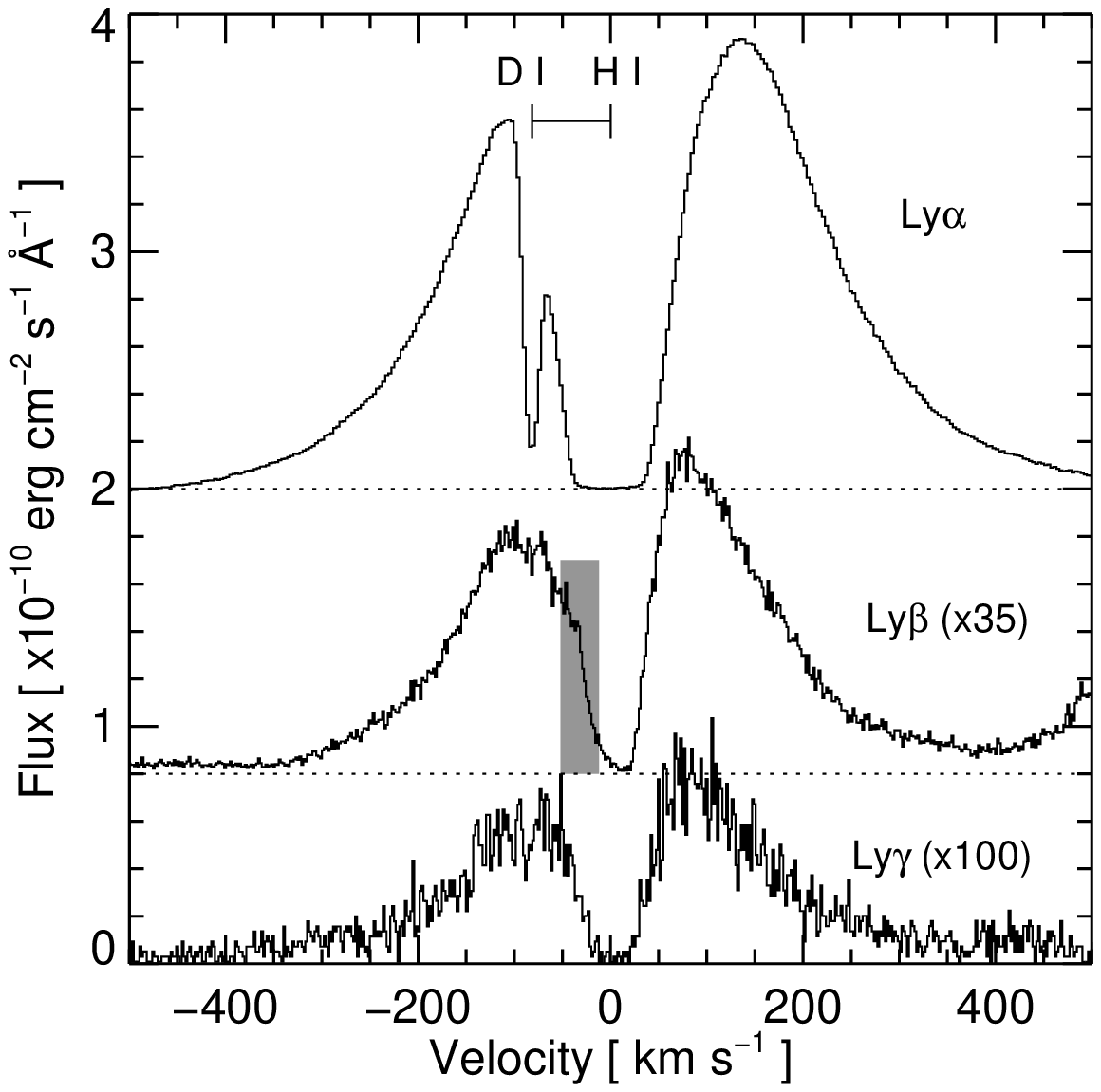}
\caption{The hydrogen Ly\gal, Ly\gb\ and Ly\ggam\ profiles for \capella. The
Ly\gal\ line is from a \emph{HST}/STIS observation on 12 Sep.\ 1999,
and the Ly\gb\ and Ly\ggam\ lines are from the MDRS LiF1A and SiC2A spectra
of \fuse\ observation P1041303. The \fuse\ spectra show only data
obtained during orbital night which minimizes airglow contamination.
However, residual airglow is still found in the Ly\gb\ profile,
indicated by the shaded region.
The Ly\gb\ and Ly\ggam\ lines have been multiplied
by the factors indicated in the figure to aid comparison. The expected
positions of the deuterium and hydrogen interstellar absorption lines
are shown.}
\label{mdrs-lybeta}
\end{figure*}

The \ion{H}{i} Ly$\alpha$ line of \capella\ has been extensively
studied in the past from \emph{Copernicus}, \emph{IUE}, and GHRS data.
A key feature is that the red peak of the line is stronger  than the
blue peak at all orbital phases (Ayres~1993), suggesting that there is
an outflow 
leading to preferential  weakening of the blue wing of the line
(Dupree~1975; Ayres~1993). The \fuse\ MDRS observation of \capella\
reveals, for 
the first time, this structure in the higher Lyman series.
A possible explanation for the
asymmetry is an acceleration throughout the chromosphere.
Fig.~\ref{mdrs-lybeta}
compares the \fuse\ \ion{H}{i} Ly\gb\ and \ion{H}{i} Ly\ggam\ lines
with the Ly$\alpha$ line obtained in Sep.\ 1999 by the Space Telescope Imaging
Spectrograph (STIS) on the \emph{HST}.

\section{Coronal Ions: Fe XVIII} 

For stars with coronae extending to temperatures of $\sim$\,$10^7$~K,
emission lines from highly-ionized ions arising from forbidden
transitions within the ground  configurations of the ions are expected
at UV wavelengths. One of the strongest is \ion{Fe}{xxi}
1354.064\,\AA, which has been observed with the GHRS  in \capella\
(Linsky \etal\ 1998), and another is \ion{Fe}{xviii}  974.85\,\AA\
which has been seen previously in solar flare spectra (Feldman \&
Doschek~1991) and is now accessible with \fuse.  Fig.~\ref{fuse-spec}
shows the presence of a weak line between \ion{H}{i} Ly\ggam\ and
\ion{C}{iii} \gl977.  By comparing both CHIANTI and APEC (Smith et
al.~2001) models of the \ion{Fe}{xviii} ion with the \fuse\  flux,
other \ion{Fe}{xviii} lines at 15.625\,\AA\ and  93.92\,\AA\ in
\emph{Chandra}\ (Canizares \etal\ 2000), and \emph{Extreme Ultraviolet
Explorer} spectra (Brickhouse \etal\ 2000), we find that the \fuse\
\gl974 flux is consistent, within a factor of 2, with the X-ray lines,
confirming the \ion{Fe}{xviii} identification.  

\begin{figure*}[h]
\figurenum{4}
\epsscale{0.5}
\plotone{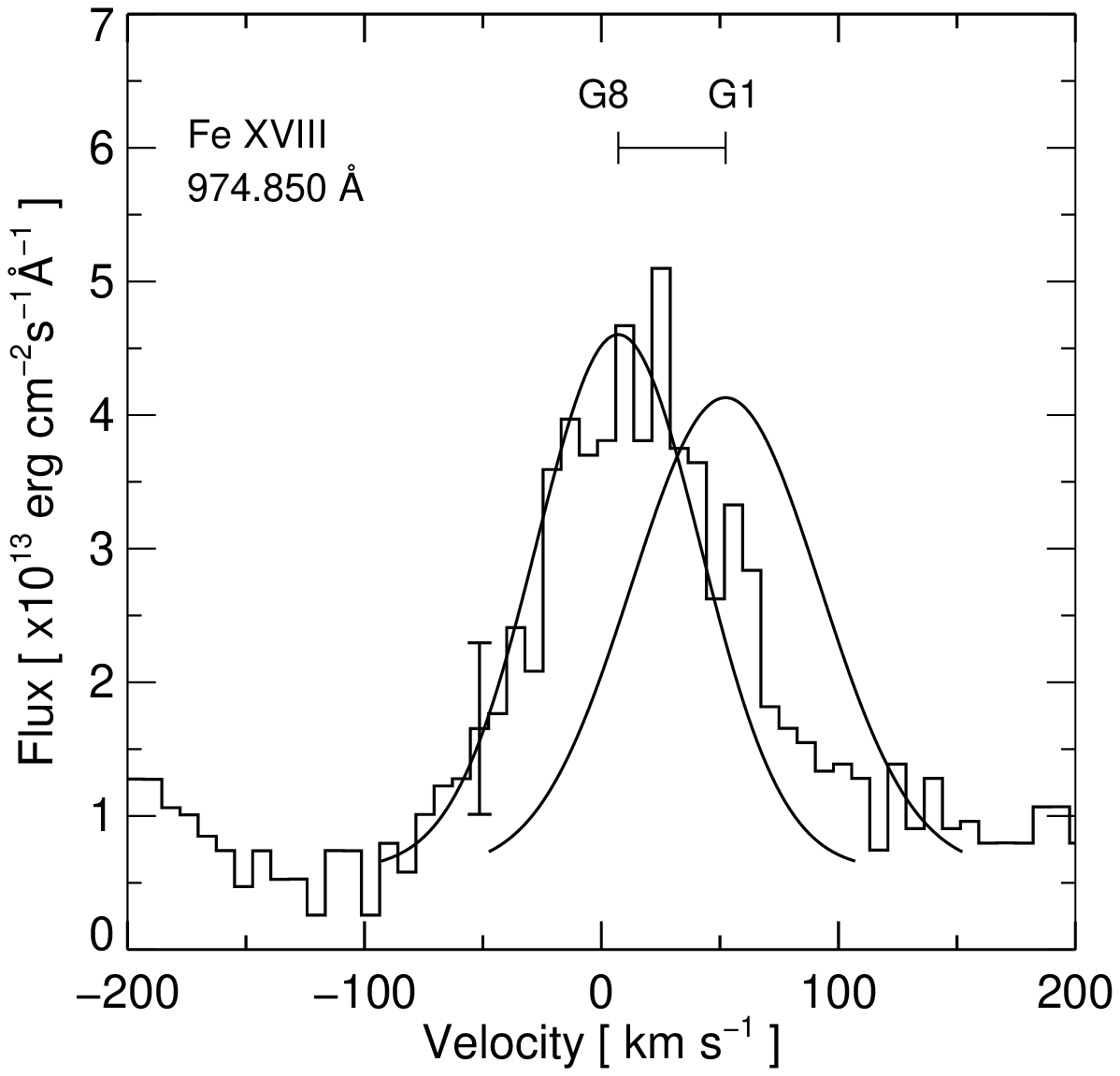}
\caption{The \ion{Fe}{xviii} \lam974 profile from the SiC2A spectrum
of observation P1041301. The zero-point of
the velocity scale corresponds to the rest wavelength of the line, 
determined from interstellar \ion{C}{iii} absorption at \gl977. The
expected velocities of the \capella\ giants are shown, together with
two profiles centered at the two stars' velocities. These profiles
are broadened with the thermal width of the \lam974 line (79~\kms),
the \fuse\ instrumental 
broadening (20~\kms), and the rotational broadenings of the two stars
(3~\kms\ for the G8 giant, and 36~\kms\ for the G1 giant), and
demonstrate the dominant contribution from the G8 giant.
The 1$\sigma$ error bar for one of the
data points indicates the signal-to-noise of the data.}
\label{fe18-plot}
\end{figure*}

The \lam974 line is narrow compared to  cooler emission lines in the
spectrum (Table 1) with a FWHM of 92 \kms. (The  thermal width of the
line is expected to be 79 \kms\ which becomes 82 \kms\ when broadened
by the instrumental profile taken as 20 \kms.)  This narrow width
indicates that the emission originates principally from only one of
the two giants.  If the nearby \ion{C}{iii} \gl977 interstellar line
(expected at a heliocentric radial velocity of $\approx$\,$+22$~\kms,
Linsky et al.~1993) is used to determine the absolute wavelength
scale, then Fig.~\ref{fe18-plot} shows that  a stronger contribution
(about 75\% of the flux) arises from the G8 giant.  To confirm this
result, the centroids of the \ion{C}{iii} \gl977 and  \ion{S}{vi}
\gl933, \gl944 lines were measured and found to be redshifted by
between 44 and 51 \kms, relative to the \ion{Fe}{xviii}
line, in the LWRS data. \ion{C}{iii} and \ion{S}{vi} would be expected
to be formed in 
the G1 atmosphere and so this confirms both the association of
\ion{Fe}{xviii} with the G8 giant and the interstellar origin of the
\ion{C}{iii} \lam977 absorption.  This \ion{Fe}{xviii} result
contrasts with a GHRS observation obtained in September 1995 where
Linsky \etal\ (1998) found almost equal contributions to the
\ion{Fe}{xxi} \lam1354 line from both stars. 
Further, the STIS spectrum obtained in Sep.\ 1999 suggests a smaller
G8 contribution to the \lam1354 line. This result together with a
discussion of these apparently conflicting GHRS,
STIS and \fuse\ results for the relative contributions of the two
giants to \capella's high temperature emission will be presented in
O.J.\ Johnson et al. (in preparation).

\section{Variability}

Monitoring of \capella\ with \emph{IUE} revealed that chromospheric
and transition region fluxes of \capella\ are variable only on the
5--10\% level (Ayres 1991).  However the higher temperature coronal
emissions observed in the EUV region are constant to $\sim$30\% except
for species formed above 6$\times$10$^6$~K which can vary by a factor
of 3 to 4 (Dupree \& Brickhouse 1995; Brickhouse \etal\ 2000).
The \fuse\ fluxes can be compared with those from
the \emph{ORFEUS I} spectrum obtained in 1993, where we find \lam977
and \lam1032 fluxes of $2.52\times 10^{-11}$ and $1.11\times
10^{-11}$~\ecs, respectively, with measurement uncertainties of around
5\%. 
These values are in reasonable agreement with the \fuse\
measurements (Table~1), although the $\approx$\,30\% difference for
the \lam1032 line is larger than the expected absolute flux accuracy
of both instruments ($\approx$\,10\%). The \lam1176 line has been
measured in both \emph{IUE} 
and GHRS spectra, with values given in Linsky et al.~(1995), and
consistency is found to within 20\%.
As the time of arrival of each photon was
recorded during the \capella\ observations,  variability can
be
sought during each observation;
however, no significant variations in the \lam977,
\lam1032 and \lam1176 fluxes were found
down to timescales of 30~s,
consistent with a lack of flaring activity. In addition no significant
change 
occurred in the flux of the \ion{Fe}{xviii} \gl974  during the 
\fuse\ observations.

\section{Summary}

The \fuse\ spectra of \capella\ illustrate the diagnostic opportunties
available for cool star research with \fuse. Line profiles and
velocities can be measured accurately, while the large number of
emission lines, many from species unavailable from \emph{IUE} or \emph{HST}
spectra, allow the temperature structure from the chromosphere
through to the upper transition region to be determined. The
\ion{Fe}{xviii} \lam974 line extends the \fuse\ temperature coverage
for active stars and provides crucial velocity information unavailable
from EUV and X-ray data.
A future paper (P.R.\ Young et al., in preparation) will discuss the
\capella\ spectrum in greater detail.

\acknowledgements

This work is based on data obtained for the Guaranteed Time
Team by the NASA-CNES-CSA \fuse\ mission operated
by Johns Hopkins University.  Financial support to US participants
has been provided by NASA contract NAS5-32985.

\begin{deluxetable}{lllll}
\tablecaption{Parameters of  Selected Emission
Lines from P1041301\label{flux-table}}
\tablehead{
Ion &Line &Flux $\times$10$^{12}$      &Photon counts  &FWHM \\
    &(\AA)&(\ecs)& (in 14174~s)     &(\kms) \\
}
\tablewidth{5in}
\startdata
\sidehead{SiC2A}
\ion{S}{vi}  &933.378  &$0.715$$^{\rm a}$  &4,980   &140 \\
\ion{He}{ii} &958$^{\rm b}$  &$0.118$  &901     &125 \\
\ion{Fe}{xviii} &974.850 &$0.113$ &848 &92 \\
\ion{C}{iii} &977.020  &$29.4$ &219,000 &247 \\
\sidehead{LiF1A}
\ion{O}{vi}  &1031.926 &$14.4$ &267,000 &162 \\
\ion{S}{iv}  &1062.664 &$0.236$ &4,490   &137 \\
\sidehead{LiF2A}
\ion{O}{i}   &1152.151 &$0.275$  &7,420    &72 \\
\ion{C}{iii} &1176$^{\rm c}$ &$15.1$ &248,00&453 \\
\enddata
\tablenotetext{a}{Includes a $\lesssim$ 10\% contribution from a
blending \ion{He}{ii} line.}
\tablenotetext{b}{A blend of seven lines between 958.670\,\AA\ and
958.725\,\AA.}
\tablenotetext{c}{A blend of six lines at wavelengths 1174.933\,\AA,
1175.263\,\AA, 1175.590\,\AA, 1175.711\,\AA, 1175.987\,\AA\ and
1176.370\,\AA.}
\end{deluxetable}

\begin{thebibliography}{}


\bibitem[]{}Ayres, T. R. 1988, ApJ, 331, 467


\bibitem[Ayres(1991)]{ayres91}
   Ayres, T. R. 1991,
   \apj\ 375, 704

\bibitem[]{}Ayres, T. R. 1993, ApJ, 402, 710

\bibitem[Barlow et al.(1993)]{barlow93}
   Barlow, D. J., Fekel, F. C., \& Scarfe, C. D. 1993,
   \pasp\ 105, 476

\bibitem[]{}Brickhouse, N. S. 1996 in IAU Colloq. 152, Astrophysics
in the Extreme Ultraviolet, ed. S. Bowyer \& R. Malina (Dordrecht:
Kluwer), 105

\bibitem[]{}Brickhouse, N. S., Dupree, A. K., Edgar, R. J.,
Liedahl, D. A., Drake, S. A., White, N. E., \& Singh, K. P.
2000, \apj\ 530, 387

\bibitem[]{}Canizares, C. R. et al. 2000, \apj\ 539, L41

\bibitem[]{}Curdt, W., Feldman, U., Laming, J. M., Wilhelm, K.,
Schuehle, U., \& Lemaire, P. 1997, A\&AS 126, 281

\bibitem[Dere et al.(2001)]{dere01}
   Dere K. P., Landi E., Young P. R., \& Del Zanna G. 2001,
   \apjs, in press

\bibitem[Doyle \& McWhirter(1980)]{doyle80}
   Doyle, J. G., \& McWhirter, R. W. P. 1980,
   \mnras\ 193, 947

\bibitem[]{}Dupree, A. K. 1975, \apj\ 200, L27

\bibitem[]{}Dupree, A. K., Foukal, P. V., \& Jordan, C. 1976,
\apj\ 209, 621

\bibitem[]{}Dupree, A. K., Brickhouse, N. S., Doschek, G. A.,
Green, J. C., \& Raymond, J. C. 1993, \apj\ 418, L41

\bibitem[]{}Dupree, A. K., \& Brickhouse, N. S. 1995, in 
IAU Symp. 176, Stellar Surface Structure, Poster
Proceedings, ed. K. G. Strassmeier, (Institut f\"ur Astronomie: Wien),
184

\bibitem[]{}
   Feldman, U., \& Doscheck, G. A. 1991,
   \apjs\ 75, 925

\bibitem[Hummel et al.(1994)]{hummel94}
   Hummel, C. A., Armstrong, J. T., \& Quirrenbach, A. 1994,
   \aj\ 107, 1859

\bibitem[]{}Kruk, J. W., Brown, T. M., Davidsen, A. F. \etal\ 1999,
\apjs\ 122, 299

\bibitem[Linsky et al.(1993)]{linsky93}
   Linsky, J. L., et al. 1993,
   \apj\ 402, 694
 
\bibitem[Linsky et al.(1995)]{linsky95}
   Linsky, J. L., Wood, B. E., Judge, P., et al. 1995,
   \apj\ 442, 381

\bibitem[Linsky et al.(1998)]{linsky98}
   Linsky, J. L., Wood, B. E., Brown, A., \& Osten, R. A. 1998,
   \apj\ 492, 767

\bibitem[Moos et al.(2000)]{moos}
   Moos, H. W., et al. 2000, \apj\ 538, L1

\bibitem[]{}Ness, J.-U., Mewe, R., Schmitt, J. H. M. M. \etal\
2001, \aap\ 367, 282


\bibitem[]{}Smith, R. K., Brickhouse, N. S., Raymond, J. C., \&
Liedahl, D. A. 2001, \apj, submitted


\bibitem[]{}Young, P. R., \& Dupree, A. K. 2001, \apj, submitted

\bibitem[Wood et al.(1997)]{wood97}
   Wood, B. E., Linsky, J. L., \& Ayres, T. R. 1997,
   \apj\ 478, 745

\end{thebibliography}
\end{document}